\documentclass{article}

\usepackage{subcaption}
\usepackage[utf8]{inputenc}
\usepackage{geometry}
\usepackage[table]{xcolor}
\usepackage{hyperref}
\usepackage{multirow}
\usepackage{graphicx}
\usepackage{comment}
\usepackage{amsmath,amssymb}
\usepackage{multicol}
\usepackage{booktabs}
\usepackage{array}
\usepackage{authblk}
\usepackage{makecell}
\usepackage[normalem]{ulem}

\renewcommand{\title}[1]{{\noindent\centering\Large\bfseries#1\medskip\\[1em]}}
\renewcommand{\author}[2]{{\noindent\large #1 \medskip\\[0.5em] \noindent \small #2 \medskip\\}}

\begin{document}

\title{Behavioral response to mobile phone evacuation alerts}

\author{E. Elejalde,\textsuperscript{1,*}
T. Naushirvanov,\textsuperscript{2}
K. Kalimeri,\textsuperscript{3}
E. Omodei,\textsuperscript{2}
M. Karsai,\textsuperscript{2,4}
L. Bravo,\textsuperscript{5} \\
L. Ferres\textsuperscript{3,5,*}
}
{
1. L3S Research Center, Leibniz University Hannover, Hannover, Germany\\
2. Department of Network and Data Science, Central European University, Vienna, Austria\\
3. ISI Foundation, Turin, Italy\\
4. National Laboratory for Health Security, HUN-REN R\'enyi Institute of Mathematics, Budapest, Hungary \\
5. IDS UDD, Santiago, Chile \\
* Corresponding: elejalde@l3s.de / lferres@udd.cl
}

\begin{abstract}
This study examines behavioral responses to mobile phone evacuation alerts during the February 2024 wildfires in Valpara\'iso, Chile. Using anonymized mobile network data from 580,000 devices, we analyze population movement following emergency SMS notifications. Results reveal three key patterns: (1) initial alerts trigger immediate evacuation responses with connectivity dropping by 80\% within 1.5 hours, while subsequent messages show diminishing effects; (2) substantial evacuation also occurs in non-warned areas, indicating potential transportation congestion; (3) socioeconomic disparities exist in evacuation timing, with high-income areas evacuating faster and showing less differentiation between warned and non-warned locations. Statistical modeling demonstrates socioeconomic variations in both evacuation decision rates and recovery patterns. These findings inform emergency communication strategies for climate-driven disasters, highlighting the need for targeted alerts, socioeconomically calibrated messaging, and staged evacuation procedures to enhance public safety during crises.
\end{abstract}

\section{Introduction}
Wildfires represent one of the most pressing environmental and societal challenges of the 21st century, driven by a combination of climate change, land-use policies, and human activities~\cite{Jones2024,guo2024global}. 
The impact of extensive wildfires like the ones experienced in California~\cite{guardian2025wildfires}
or the Amazon rain-forest~\cite{cursino2024valparaiso}
goes beyond the local scale, affecting climate patterns, human health, and the economy~\cite{burke_exposures_2022,mcconnell2024rare,wangSevereGlobalEnvironmental2024}.
While the evacuation procedures of wildfires are well-documented~\cite{thompson_evacuation_2017}, the role of near-real-time communication technologies in shaping human responses to these crises remains less understood. 
Rapid and structured evacuation is critical to minimize loss of life during wildfires, yet human decision-making in such high-stress situations is influenced by multiple factors, including message timing, perceived risk, and socioeconomic status.

The widespread penetration of mobile technology has transformed emergency alert systems into essential tools for crisis management, enabling the rapid dissemination of evacuation directives, curfews, and updates on available assistance~\cite{ismagilova2019smart}. As of 2024, there are 112 mobile-cellular subscriptions per 100 inhabitants~\cite{ITU2024} worldwide, and Chile is in line with this trend, reporting 133 phones per 100 inhabitants as of June 2024~\cite{subtel2024mobile}. 
This ubiquity has positioned Short Message Service (SMS) notifications as a viable and widely adopted method for mass communication during crises.
A key advantage of SMS-based alerts is their ability to function on basic mobile phones without requiring Internet connectivity. Moreover, SMS is particularly effective in reaching diverse populations, including those in remote or low-resource settings, ensuring timely access to critical information.
Building on the effectiveness of SMS, the Wireless Emergency Alerts (WEA) system was developed to send geo-targeted emergency messages to mobile devices by authorized public safety officials~\cite{federal_communications_commission_wireless_2024}. 
WEAs, operating via cell broadcasting, deliver warnings within seconds, without the need for a subscription or internet connection~\cite{cisa2024wea}.
By integrating SMS capabilities with advanced alerting systems like WEA, governments have enhanced emergency communications, ensuring rapid, reliable, and inclusive public warnings in times of crisis~\cite{haddow2013disaster}.
These systems have demonstrated critical in saving lives and managing emergencies effectively.
One of the earliest recorded cases of SMS for emergency alerts dates back to 19 September 2007, when the Disaster Management Center of Sri Lanka sent a 20-word text alert following a magnitude 7.9 earthquake off the southern coast of Sumatra~\cite{columbia2007sms}.
During the 2010 Haiti earthquake, humanitarian organizations used SMS to coordinate rescue efforts and provide vital information to affected people \cite{coyle_new_2009}. During the outbreaks of Ebola in Liberia and Sierra Leone, SMS was used to spread public health messages and combat misinformation, significantly aiding disease control efforts \cite{danquah_use_2019, berman_use_2017, trad_guiding_2015} and even controlling the spread of disease-associated rumors \cite{turner_rumour_2023}. In Japan, the J-Alert National Early Warning System serves as the country's primary platform for rapidly disseminating critical information to the public \cite{wikipedia2025jalert}.

Despite the demonstrated utility of SMS and alert systems in crisis management~\cite{cinnamon_evidence_2016}, their effectiveness in prompting protective action remains an open question. 
To date, the main evaluation of the messaging effectiveness relied on surveys ~\cite{carlson2024360} and agent-based models~\cite{gao2022household}.
Such approaches offer valuable insights, still, are prone to self-reporting biases regarding actual human behavior during disasters, which is shown to be influenced by the timing, content, and perceived credibility of risk communication~\cite{slovic1987perception, durage2014evacuation,  kuligowski2021evacuation}. 
The main reason stems primarily from the lack of accessible, high-resolution data that can act as a proxy to assess near-real-time evacuation patterns.
Hence, gaps persist in understanding how individuals interpret and respond to emergency messages in real-world wildfire events. 
Do initial alerts trigger immediate evacuation, or does repeated exposure to messages induce alert fatigue, leading to reduced responsiveness? Are individuals from different socioeconomic backgrounds equally likely to heed evacuation warnings, or do structural inequalities create disparities in mobility and safety?

The physical devastation of the wildfires that erupted at Valpara\'iso in Chile on February 2nd, 2024, was the most severe in the country over the last three decades. It caused 137 deaths, made 1,600 others homeless, and directly affected more than 16,000 people~\cite{cursino2024valparaiso}, with the Chilean Government, as a consequence, declaring a State of Emergency and Catastrophe. 
Chile has also implemented a similar SMS-based warning system, the \textit{Sistema de Alerta de Emergencia} (SAE), following the 27 February 2010 earthquake, to enhance public warnings for natural disasters. Since March 2017, all mobile phones sold in Chile have supported SAE at no additional cost to users, ensuring accessibility regardless of phone balance. From 2019 onward, mobile devices in Chile have come with SAE pre-configured by default~\cite{senapred2024alerts}.
This study leverages anonymized mobile network data to examine population-wide behavioral responses to evacuation alerts. 
More specifically, our study addresses three key research questions regarding behavioral responses, effectiveness, and unintended consequences of emergency messaging:
First, \textit{how does the timing and sequence of emergency alerts influence evacuation behavior?} 
Initial alerts may prompt a stronger response due to their urgency and element of surprise, while follow-up messages may be perceived as repetitive or less critical, potentially leading to alert fatigue and diminished compliance. Understanding this dynamic is crucial for optimizing the timing, frequency, and content of alerts to maintain public engagement without causing desensitization~\cite{dash2007evacuation,slovic1987perception}.
Second, \textit{to what extent do socioeconomic factors shape evacuation patterns?} Socioeconomic status influences access to evacuation resources, comprehension of emergency messages, and trust in authorities~\cite{bonfanti2023role,thompson_evacuation_2017}. Higher socioeconomic groups (SEG) of individuals may have better access to transportation and multiple information sources, enabling quicker and more independent responses. In contrast, lower SEG may face barriers such as limited mobility and historical mistrust of government-issued alerts, which could hinder effective evacuation. Addressing these disparities is essential for ensuring equitable crisis response strategies~\cite{yabe2022mobile}. 
Third, \textit{what are the unintended consequences of non-targeted alerts on population movement?} Broadly disseminated emergency messages can trigger unnecessary evacuation in areas not directly threatened by wildfires, potentially straining transportation infrastructure and emergency resources, diverting them from areas in need~\cite{yabe2022mobile}, while they can also have political implications~\cite{calderara_union_2024}.
 
Our study contributes to the growing literature on human adaptation to environmental hazards by integrating high-resolution behavioral data with policy-relevant questions. As climate change exacerbates wildfire risk globally, optimizing emergency communication strategies will be critical to enhancing public safety and resilience. 

\section{Results}

\begin{figure}[h]
    \centering
    \begin{subfigure}[t]{0.33\linewidth}
        \centering
        \includegraphics[width=\linewidth]{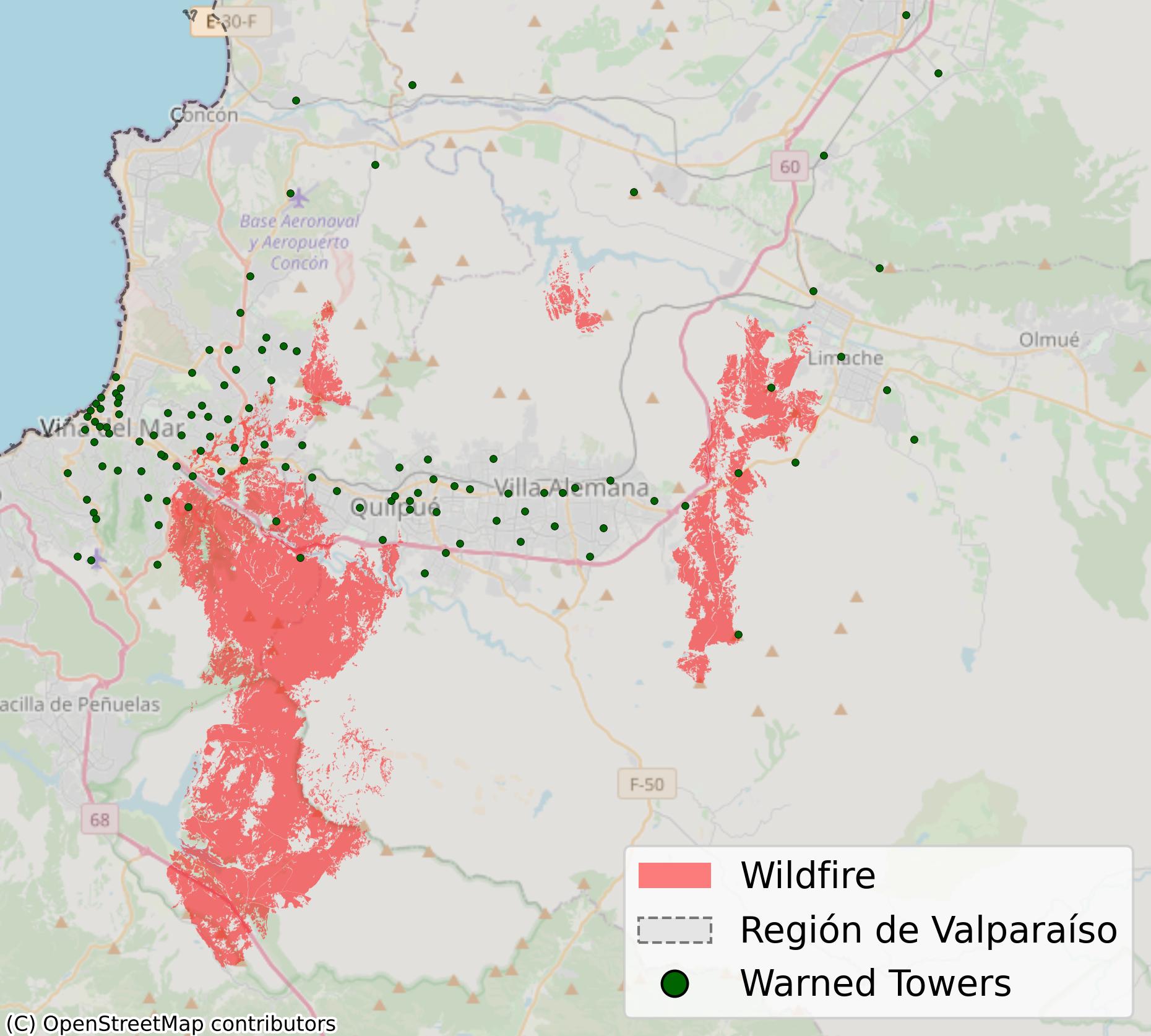}
    \end{subfigure}
    \begin{subfigure}[t]{0.66\linewidth}
        \centering
        \includegraphics[width=\linewidth]{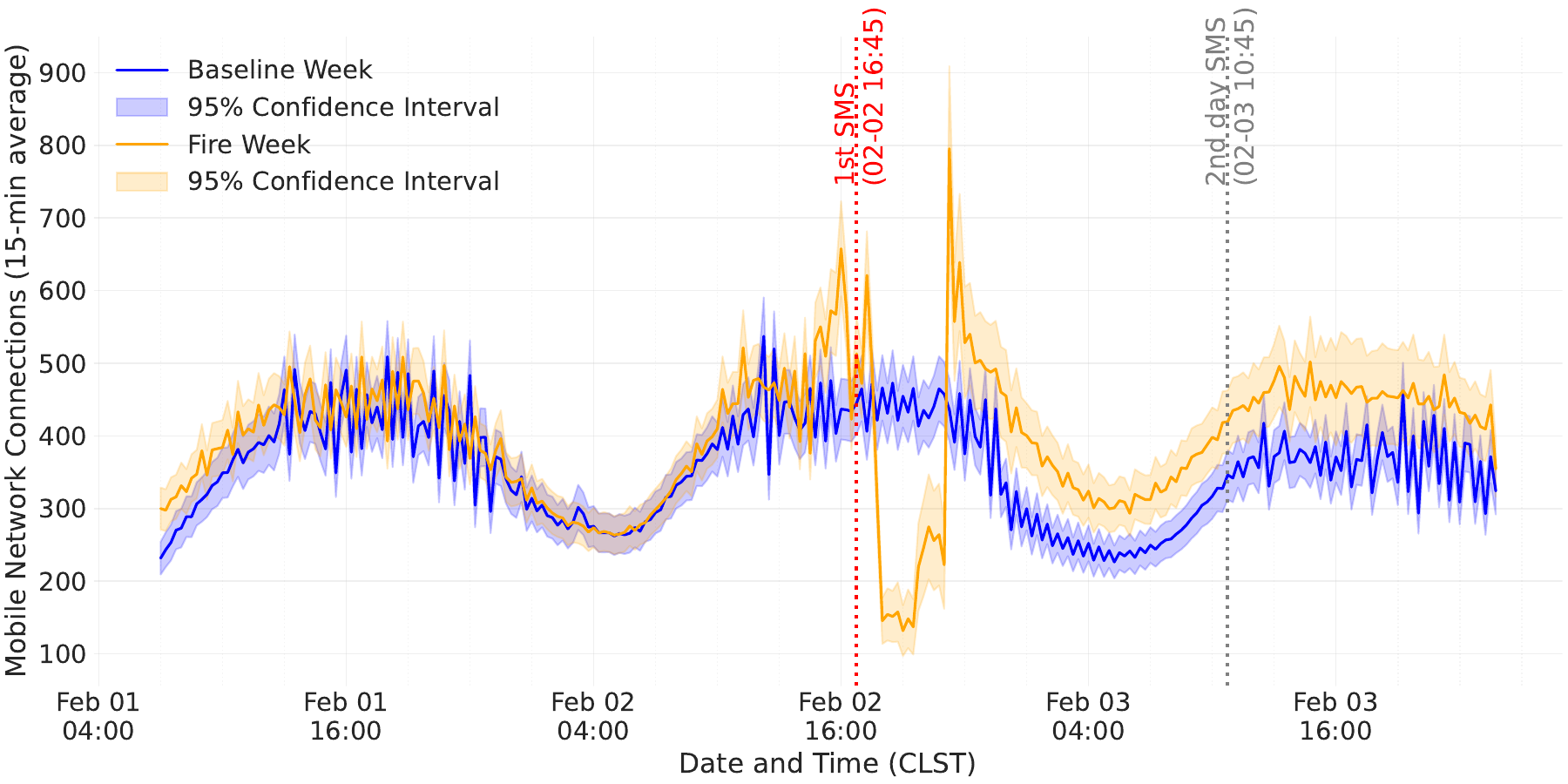}
        \label{fig:timeseries}
    \end{subfigure}%

    \caption{\textbf{Mobile phone tower connectivity patterns and location during a wildfire evacuation event.} \textbf{A}. General geographic area of the fires (in red) along with the warned towers (green dots). \textbf{B}. Time series comparing average mobile tower connections during baseline days (blue) and fire week (orange) in February 2024, with 95\% confidence intervals (shaded areas). Vertical lines indicate critical emergency communication times: first evacuation alert (red, February 2, 16:45) and first alert of day two (gray, February 3, 10:45). The data reveals distinct behavioral patterns following the initial evacuation alert, including an immediate spike in connectivity followed by a rapid decrease, suggesting population displacement.  
    }
    \label{fig:evacuation_analysis}
\end{figure}

\begin{figure}[t]
    \centering
    \includegraphics[width=1\linewidth]{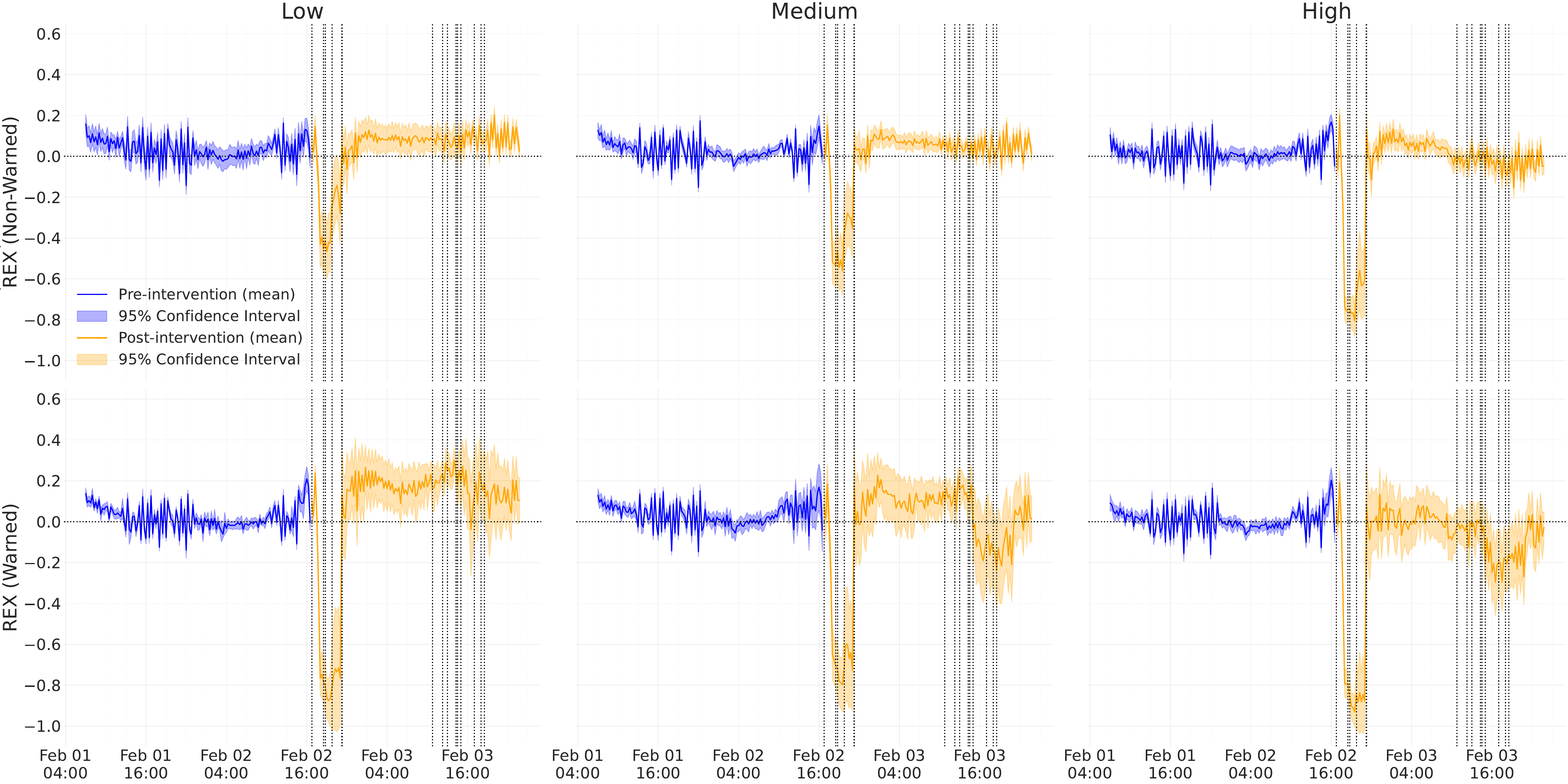}
    \caption{\textbf{Relative Evacuation Index for non-warned  (upper panel) and warned towers (lower panel), across SEGs before and after the first evacuation alert}. The blue lines represent pre-warning periods, while the orange lines indicate post-warning periods. Vertical dashed lines correspond to the times when evacuation alerts were sent. Panels are divided by SEG (low, medium, high) and whether towers were warned or non-warned. The shaded regions indicate 95\% confidence intervals. The figure illustrates behavioral differences in evacuation responses and recovery patterns across SEGs and between warned and non-warned areas.}
    \label{fig:diff_norm_no_align}
\end{figure}

On February 2, 2024, at 16:45, the first official evacuation alert was issued to towers in the Valparaíso region (Table \ref{tab:message_times} shows the analyzed messages). The Valparaíso region (Figure \ref{fig:evacuation_analysis}A) lies along the central coast of Chile between approximately 32°02' and 33°57' South latitude and 70°00' and 71°43' West longitude. The region covers an area of 16,396 square kilometers and is characterized by diverse topography, including coastal plains, the Coastal mountain range, and portions of the Andes mountains. The region has a Mediterranean climate with dry summers, which makes it prone to wildfires, and mild, wet winters. It houses a population of roughly 1.8 million inhabitants, with the majority concentrated in the Greater Valparaíso metropolitan area. The region's capital, Valparaíso city, is a UNESCO World Heritage site located approximately 120 kilometers northwest of Santiago, Chile's capital. 
We analyzed anonymized phone network data aggregated over 15-minute intervals; Figure \ref{fig:evacuation_analysis}B depicts the near-real-time device activity in reference to the baseline period (see Section \ref{sec:methods}).   
We notice that prior to the first evacuation alert, the connectivity patterns--that is, the number of unique active mobile phones in the different towers--aligned closely with baseline days measurements with the connectivity to follow the expected circadian pattern with more connections and variability during day hours~\cite{hazarie2020uncovering}. 
However, an anomaly is observed right before and after the first alert, indicating a deviation from the normal connectivity patterns.
This response evolves in three distinct phases.
The initial phase, preceding the evacuation alert, exhibits an increase in activity that lasted until shortly after the first alert. This phase shows several spikes to a maximum average of approximately 658 (95\% CI 592, 723) connections—for a 50\% increase on the baseline average. This surge suggests heightened mobile phone usage outside WiFi coverage, possibly driven by information-seeking behavior and coordination efforts.  A second phase followed with a large drop, sinking to 132 connections on average (95\% CI 97, 167), indicating rapid population displacement from the area. 
Despite multiple subsequent alerts (see Figure \ref{fig:diff_norm_no_align}), connectivity levels gradually stabilized to pre-alert values over an 8-hour period, returning to a circadian pattern around 00:00-01:00 on February 3 and starting a third phase of recovery. 
The second day (recovery phase) exhibited sustained elevated activity 40-20\% above typical patterns, with altered temporal signatures persisting throughout the observation period. This suggests enduring modifications to local connectivity patterns following the emergency event. 
Notably, the first alert of the second day showed no visible effect on connectivity patterns, unlike the response to the initial evacuation message, highlighting the diminishing effectiveness of repeated alerts and potential alert fatigue.

\begin{figure}[t]
    \centering
    \includegraphics[width=1\linewidth]{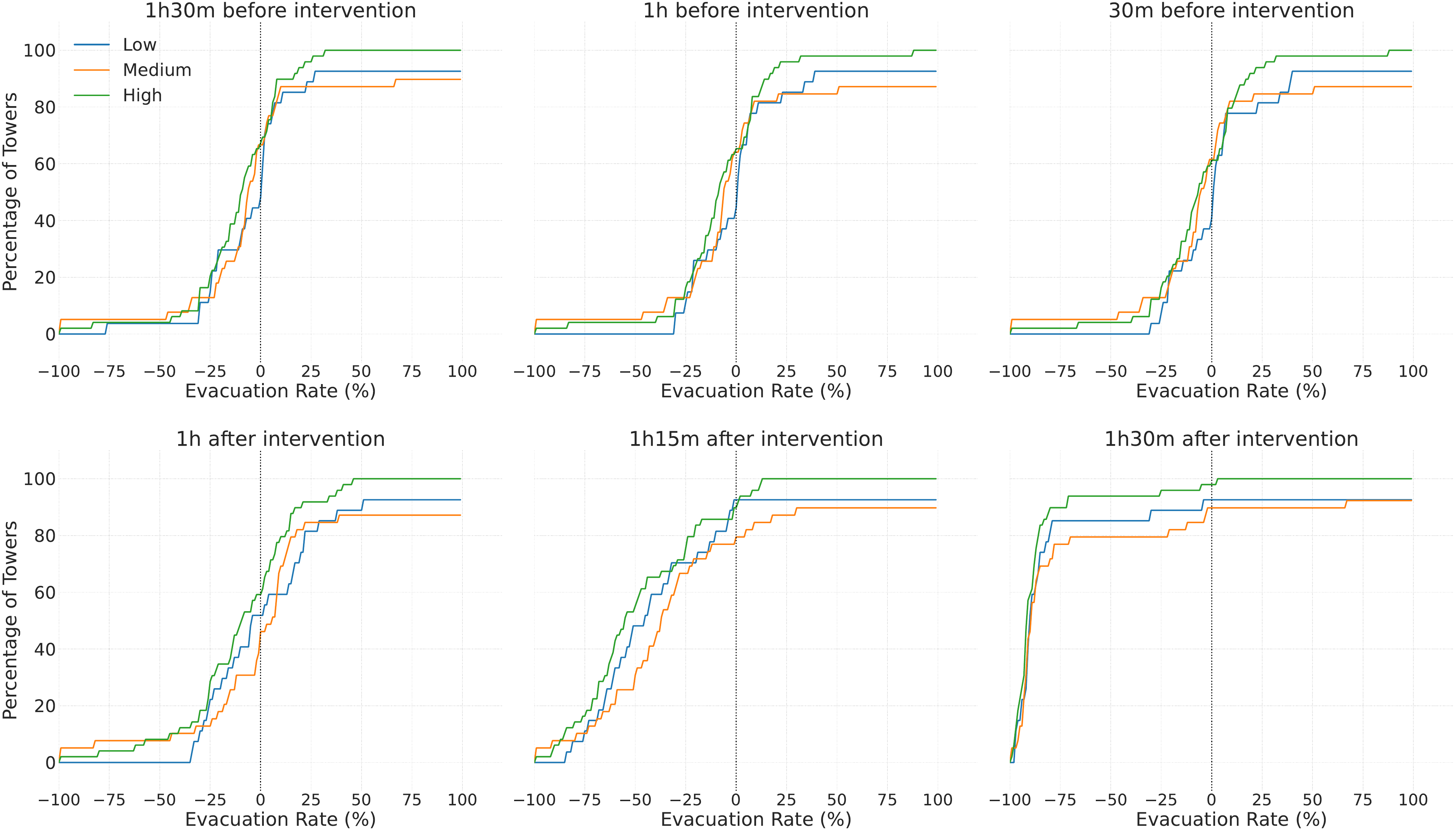}
    \caption{\textbf{Snapshots of the Cumulative distribution functions (CDFs) of tower-level evacuation rates with respect to the baseline behavior, stratified by socioeconomic status}. The figure presents six time-sliced CDFs capturing activity patterns before and after the intervention across low (blue), medium (orange), and high (green) socioeconomic areas. The top row of the panel shows the pre-intervention behavior at 90 minutes,  1 hour, and 30 minutes, respectively, while the second row in the panel indicates the post-intervention behavior displaying changes at 1 hour, 75 minutes, and 90 minutes after the intervention. The x-axis represents the percentage of the evacuation rate, where the ``0'' rate indicates a consistent behavior with the baseline days. The negative rates indicate an evacuation behavior with respect to the baseline, while the positive rates represent more connections with respect to the baseline. The y-axis expresses the cumulative percentage of towers exhibiting evacuation rates lower or equal to the corresponding evacuation rate in time (x-axis).}
    \label{fig:CDFs_diff_first_ever}
\end{figure}

Diving into the socioeconomic disparities, we stratify by socioeconomic groups (see Supplementary Figure \ref{suptbl:supp_rawsocioecon}), and differences become apparent, particularly during the recovery phase. High-income areas show less variability in the average number of connections and return to baseline ranges faster. This suggests that populations in wealthier regions experience fewer barriers to evacuation and recovery, such as alternative housing options.
The cumulative distribution functions of tower-level evacuation rates relative to baseline behavior (see Figure \ref{fig:CDFs_diff_first_ever}) further reveal distinct temporal patterns in device activity across the three socioeconomic groups (low, medium, and high) before and after the alert intervention. In the pre-intervention period (1 hour 30 minutes to 30 minutes to intervention), activity patterns remained broadly consistent with baseline behavior, as indicated by the clustering of CDFs around the zero point of the evacuation rate. The figure shows that most towers stay within $\pm$25\% of the baseline during this time. However, a marked divergence emerged post-intervention, particularly at 1 hour and 15 minutes since the alert. At this time point, approximately 60\% of towers in high socioeconomic areas showed a 50\% evacuation rate, compared to roughly 50\% in low and less than 30\% in medium socioeconomic areas. The separation between socioeconomic groups became most pronounced at 1 hour 30 minutes post-intervention, where high socioeconomic areas exhibited the strongest evacuation response, with nearly 90\% of towers showing at least 75\% evacuation. In contrast, in low and medium areas, only 85\% and 77\% of the towers reach the same evacuation rate, respectively. The observed response across socioeconomic categories suggests that the alert intervention triggered an unequal evacuation behavior, with populations in higher socioeconomic areas demonstrating substantially greater activity changes in warned towers. Notably, the steepness of the CDF curves in the post-intervention period, particularly between  1 hour 15 minutes and  1 hour 30 minutes, indicates a relatively uniform response within each socioeconomic group, suggesting coordinated population movement patterns that reflect underlying social and economic disparities in evacuation capacity.

\begin{figure}[t]
    \centering
    \includegraphics[width=\linewidth]{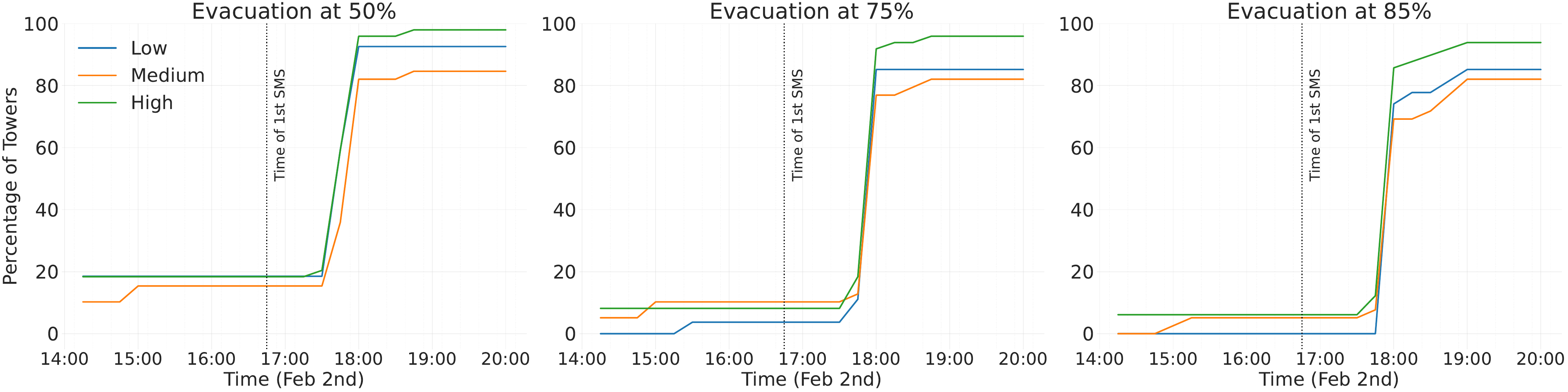}
    \caption{\textbf{Cumulative distribution functions of tower evacuations over time}. The y-axis indicates the percentage of towers that reached the specified evacuation rates (50\%, 75\%, and 85\%) at a specific time (x-axis). We stratified by socioeconomic groups: low (blue line), medium (orange line), and high (green line). The dashed line marks the time at which the first evacuation alert was sent.}
    \label{fig:CDFs_time_first_ever}
\end{figure}

To zoom in on the evacuation temporal patterns and the delay in response to the first system-wide evacuation alert, we also plot the cumulative distribution over time for the percentage of towers at different evacuation rates (50\%, 75\%, and 85\%) (Figure \ref{fig:CDFs_time_first_ever}), where 100\% is complete evacuation, meaning towers had no connections. Prior to the alert (marked by the vertical dashed line at 02-02 16:45), all socioeconomic groups showed minimal evacuation activity between 14:00 and 16:45. Following the evacuation alert, there was a dramatic and nearly simultaneous response across all groups. Among all evacuation rates, high socioeconomic areas demonstrated the most vigorous response, with over 95\% of their towers showing at least 50\% evacuation within 1 hour and 15 minutes of the alert (by 02-02 18:00), compared to 92\% for low and 82\% for medium socioeconomic areas. The socioeconomic gradient became more pronounced at higher evacuation rates. At the 75\% evacuation level, while high socioeconomic areas maintained nearly 92\% compliance by 18:00, medium socioeconomic areas showed a reduced response of approximately 77\% of towers, and low socioeconomic areas dropped to about 85\%. The disparity is most stark at the 85\% evacuation rate, where by 18:00, about 86\% of towers in high socioeconomic areas, 69\% in medium areas, and 74\% in low socioeconomic areas achieved this evacuation level. These patterns suggest that while the alert triggered an immediate response within the first 1 hour and 15 minutes across all socioeconomic groups, the capacity to achieve higher levels of evacuation was strongly influenced by socioeconomic status.

The temporal CDF plots show that low SEG towers were slower to start with the evacuation process at higher rates (e.g., reaching evacuation rates above 75\%) than medium SEG towers. However, after 1 hour and 15 minutes, they reacted, and by 1 hour and 30 minutes after the first evacuation alert, approximately 85\% of the low SEG towers had reached a 75\% evacuation rate, overcoming the 77\% of medium SEG towers at the same evacuation rate. This indicates a higher impact of the event on the lower SEG while maintaining a longer reaction time.
The high SEG towers are the fastest in reaction time and most compliant with evacuation. The figures show that almost 95\% of high SEG towers reached over 75\% evacuation rate within 1 hour and 30 minutes after the first system-wide alert. This reveals their higher reaction capacity to emergency events. However, when considering the behavior of non-affected towers, the pattern observed for high SEG areas may also point to an overreaction to the warning alerts.

To facilitate a standardized comparison across heterogeneous spatial units, we introduce the Relative Evacuation Index (REX), a normalized metric for quantifying population displacement, defined as: 

$$
\text{REX} = \frac{n_t - n_b}{n_t + n_b}
$$

\noindent where $n_t$ represents the number of unique mobile devices detected in a given spatiotemporal unit at time $t$, and $n_b$ denotes the device count for the corresponding spatiotemporal unit in the baseline days. This formulation maps the relative population change onto the interval [-1, 1]. REX exhibits several desirable methodological advantages: it maintains symmetry around zero, ensures scale invariance across different population densities, and exhibits stability compared to conventional percentage-based measures, particularly when baseline populations are small. The resulting index provides an intuitive interpretation where negative values indicate net population egress, positive values suggest population ingress, and the magnitude reflects the relative strength of displacement. Supplementary figures \ref{fig:CDFs_diff_first_ever_norm} and \ref{fig:CDFs_time_first_ever_norm} show the cumulative distribution functions using REX instead of the evacuation rate (cf. Figures \ref{fig:CDFs_diff_first_ever} and \ref{fig:CDFs_time_first_ever} above).

We find that both warned and non-warned towers reacted strongly to the evacuation alert (see Figure \ref{fig:diff_norm_no_align}). In non-warned towers, the REX reached -0.47 (95\% CI, -0.60, -0.33) for the low SEG, -0.56 (95\% CI, -0.68, -0.44) for the medium SEG and -0.81 (95\% CI, -0.88, -0.73) for the high SEG. This indicates that many people chose to leave the area regardless of whether they received the message addressing them directly or not. These findings have important policy implications, as individuals not directly affected by the warning may still attempt to evacuate, potentially congesting main roads and impacting those required to evacuate. While non-warned areas show evacuation behavior in general, the response is stronger among towers in the high SEG compared to those in the low and medium SEGs. Specifically, while the evacuation rate approximately doubles for low -0.87 (95\% CI, -0.98, -0.77) and medium -0.80 (95\% CI, -0.93, -0.66) SEGs when their communities are directly addressed (warned towers), the difference between warned and non-warned towers in the high SEG is only around 15\%, increasing from -0.81 in non-warned towers to -0.93 (95\% CI, -1.00, -0.86) in warned towers. As observed before, SEG differences do not significantly affect the time required for these areas to reach the lower REX (all between 18:30 and 19:30). However, they do influence the intensity of the effect and behavior during the recovery phase between the last message on February 2 and the end of the observed period.

\begin{figure}
    \centering
    \includegraphics[width=1\linewidth]{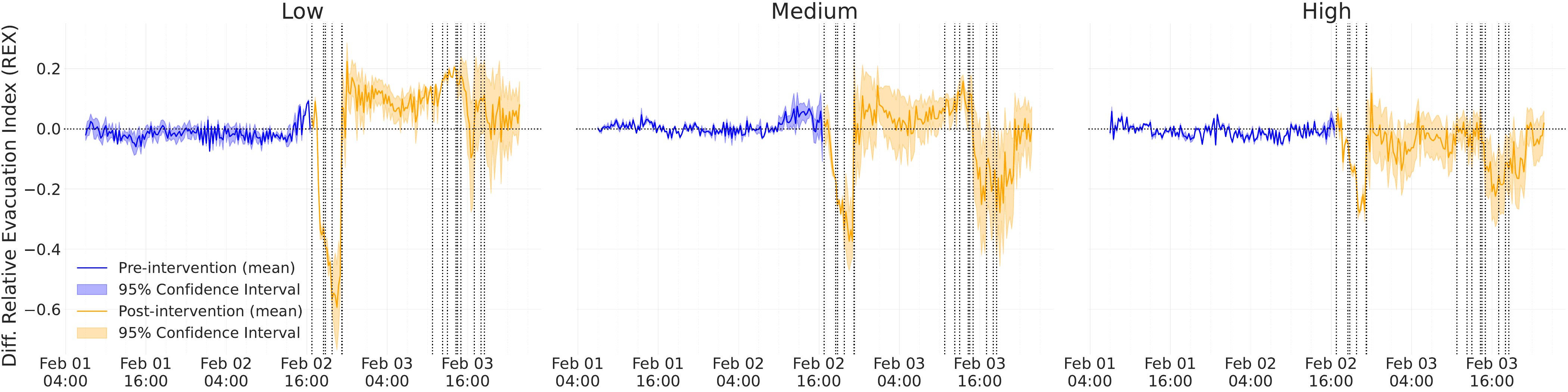}
    \caption{\textbf{Differential REX over time stratified by three socio-economic groups (Low (left), Medium (center), and High (right)) from February 1-3, 2024}. We depict the pre-intervention (blue) and post-intervention (orange) time series, with their respective 95\% confidence intervals, and measurements taken every 15 minutes. The dashed lines are all the evacuation alerts sent.}
    \label{fig:diff_diff_norm_no_align}
\end{figure}

High SEG evacuees return to normal levels faster than those in the low and medium SEGs, practically skipping the recovery phase observed in other groups. Low SEG evacuees maintain a stable 20\% increase over the baseline, while medium SEG evacuees stabilize at around 10\%. A notable effect is observed in response to the second batch of evacuation messages on February 3: high SEG shows a stronger reaction again, reaching a mean REX of -0.25. The medium SEG reaches -0.20, though this is not a statistically significant deviation from the baseline at the 95\% CI, while the low SEG does not show a noticeable change and remains above the baseline. As observed before, the REX measurements show that warning messages sent during the second day do not prompt the same reaction seen for the initial SMSs on the previous day, even for communities that were first notified on February 3rd (see Figures \ref{fig:rex-first-day} and \ref{fig:rex-second-day}).
Note that, unlike on the first day, the differentiated reaction during the second day for the high SEG might be explained, at least in part, by the distribution of the affected areas, where more than half of the notified towers belong to the high SEG.
To isolate the intervention's true impact across SEGs, we used non-affected towers as a baseline control. This approach allowed us to distinguish intervention-specific responses from general panic-motivated behavior. Notably, while affected areas showed distinct response patterns, non-affected towers quickly returned to pre-intervention activity levels without a recovery phase. The difference in REX between affected and non-affected towers reveals clear socioeconomic patterns (Figure \ref{fig:diff_diff_norm_no_align}). During the pre-intervention period, no SEG showed significant behavioral deviations from the previous week. However, following the first evacuation alert, the magnitude of the adjusted REX decline correlated directly with the tower locations' socioeconomic status, with the strongest effects observed in low-income areas. In contrast, high-SEG communities showed minimal changes in the adjusted REX, suggesting these populations would likely have evacuated regardless of whether they were in affected areas.

\begin{table}[ht]
\centering\scriptsize
\caption{Controlled Interrupted Time Series (CITS) Model for different socioeconomic groups (SEG).}
\label{tab:regression-2days}
\begin{tabular}{lccc|ccc}
\toprule\toprule
& \multicolumn{3}{c}{Socio-Economic Groups (Feb 2nd)} & \multicolumn{3}{c}{Socio-Economic Groups (Feb 3rd)}\\
\cmidrule(lr){2-7}
& Low & Medium & High & Low & Medium & High \\
 &  &  &  &  &  &  \\
\midrule
N & 8143 & 10189 & 15804 & 8283 & 10018 & 15588 \\
$R^2$ & 0.289 & 0.329 & 0.427 & 0.025 & 0.026 & 0.032 \\
\midrule
Intercept & \makecell{-0.051***\\(-0.081--0.022)} & \makecell{-0.045***\\(-0.067--0.023)} & \makecell{-0.053***\\(-0.075--0.032)} & \makecell{0.141*\\(0.012-0.271)} & \makecell{0.256***\\(0.179-0.333)} & \makecell{0.487***\\(0.399-0.576)} \\
$T$ & \makecell{0.001***\\(0.0-0.001)} & \makecell{0.001***\\(0.0-0.001)} & \makecell{0.001***\\(0.0-0.001)} & \makecell{0.0\\(-0.001-0.0)} & \makecell{-0.001***\\(-0.001--0.001)} & \makecell{-0.002***\\(-0.003--0.002)} \\
$I_0$ & \makecell{-0.393***\\(-0.433--0.353)} & \makecell{-0.463***\\(-0.498--0.428)} & \makecell{-0.675***\\(-0.7--0.651)} & \makecell{-0.002\\(-0.028-0.024)} & \makecell{-0.003\\(-0.018-0.011)} & \makecell{0.005\\(-0.014-0.023)} \\
$T_{I_0}$ & \makecell{0.026***\\(0.023-0.029)} & \makecell{0.031***\\(0.029-0.034)} & \makecell{0.045***\\(0.044-0.047)} & \makecell{0.001\\(-0.0-0.002)} & \makecell{0.001***\\(0.001-0.002)} & \makecell{0.002***\\(0.001-0.002)} \\
$G$ & \makecell{-0.065**\\(-0.107--0.022)} & \makecell{-0.054*\\(-0.099--0.01)} & \makecell{-0.042*\\(-0.077--0.007)} & \makecell{0.128\\(-0.124-0.381)} & \makecell{0.0\\(-0.226-0.226)} & \makecell{-0.228\\(-0.467-0.01)} \\
$GT$ & \makecell{0.0*\\(0.0-0.001)} & \makecell{0.001*\\(0.0-0.001)} & \makecell{0.0\\(0.0-0.001)} & \makecell{0.0\\(-0.002-0.001)} & \makecell{0.0\\(-0.001-0.001)} & \makecell{0.001\\(0.0-0.002)} \\
$GI_0$ & \makecell{-0.36***\\(-0.421--0.3)} & \makecell{-0.242***\\(-0.302--0.182)} & \makecell{-0.114***\\(-0.156--0.073)} & \makecell{0.054*\\(0.004-0.104)} & \makecell{-0.046*\\(-0.091--0.002)} & \makecell{-0.073**\\(-0.121--0.025)} \\
$GT_{I_0}$ & \makecell{0.03***\\(0.025-0.036)} & \makecell{0.015***\\(0.01-0.02)} & \makecell{0.005*\\(0.001-0.009)} & \makecell{-0.003***\\(-0.005--0.001)} & \makecell{-0.004***\\(-0.005--0.002)} & \makecell{-0.001\\(-0.003-0.001)} \\
\midrule\midrule
\multicolumn{6}{@{}l@{}}{\footnotesize Note: $^{*}p<0.05$; $^{**}p<0.01$; $^{***}p<0.001$}
\end{tabular}
\end{table}

We used a controlled interrupted time series (CITS) analysis to evaluate the causal effect of the warning alert intervention during the initial 24-hour period (see Section \ref{sec:methods} below). Analysis of the pre-intervention temporal trends revealed no significant association between the REX and time for either the treatment or control groups ($\beta_1 \approx 0$ and $\beta_5 \approx 0$, respectively). The statistical models demonstrated that SEGs maintained consistent connectivity patterns during the pre-intervention period relative to the baseline days, indicating stability in connectivity behavior prior to the intervention. Also, analysis of baseline differentials ($G$ coefficient) revealed minimal variations across SEGs, with relatively small magnitude differences ($\delta < 0.1$). These findings suggest negligible pre-existing heterogeneity in evacuation behaviors across different socioeconomic groups prior to the intervention, strengthening the validity of our causal analyses.

Examination of immediate intervention effects revealed substantial negative coefficients ($I_0$) across all SEGs in the control condition, indicating rapid evacuation behavior in response to the initial evacuation warning. The magnitude of this effect exhibited a positive correlation with socioeconomic status ($I_0^{Low} = -0.393$, $I_0^{Med} = -0.463$, $I_0^{High} = -0.675$), suggesting heightened reactivity to warning messages among higher SEGs. The group-intervention interaction term ($GI_0$) demonstrated additional negative effects that were inversely proportional to socioeconomic status ($GI_0^{Low} = -0.360$, $GI_0^{Med} = -0.242$, $GI_0^{High} = -0.114$). This pattern indicates that the evacuation mandate had a more pronounced impact on lower SEGs, suggesting these groups may be less inclined to evacuate without explicit governmental directives.

Post-intervention return patterns revealed positive trend changes ($T_{I_0}$) across all groups, with the majority of tower locations showing population recovery within 12 hours of evacuation. The control group demonstrated a socioeconomic gradient in return rates, with higher SEGs exhibiting stronger coefficients ($T_{I_0}^{High} = 0.045$, $T_{I_0}^{Med} = 0.031$, $T_{I_0}^{Low} = 0.026$), corresponding to approximately 5\% REX recovery per 15-minute interval in high SEG areas. The group-trend interaction terms ($GT_{I_0}$) exhibited additional positive effects that were inversely proportional to socioeconomic status ($GT_{I_0}^{Low} = 0.030$, with diminishing magnitudes as SEG increased), indicating accelerated population return rates among lower SEGs in affected communities. Note that a faster return rate does not necessarily mean a faster recovery to pre-intervention values, as the lower SEGs exhibit a stable but sustained pattern of hyperactivity during the entire day of Feb 3rd.  
Nevertheless, the CITS models for February 3rd show that the first warning message of the second day did not have the same mobilizing effect as in the previous day. In this case, we do not see a significant impact for not affected towers either immediately ($I_0$) or in the longer trend ($T_{I_0}$), thus suggesting that these communities had already overcome the initial rush of the event. Also, the observed impact for the affected towers is one order of magnitude smaller than the day before (see also Figure \ref{fig:model_day2}). These results support the hypothesis of a diminished effect / alert fatigue for subsequent rounds of messages during a critical event.

\begin{figure}
    \centering
    \includegraphics[width=1\linewidth]{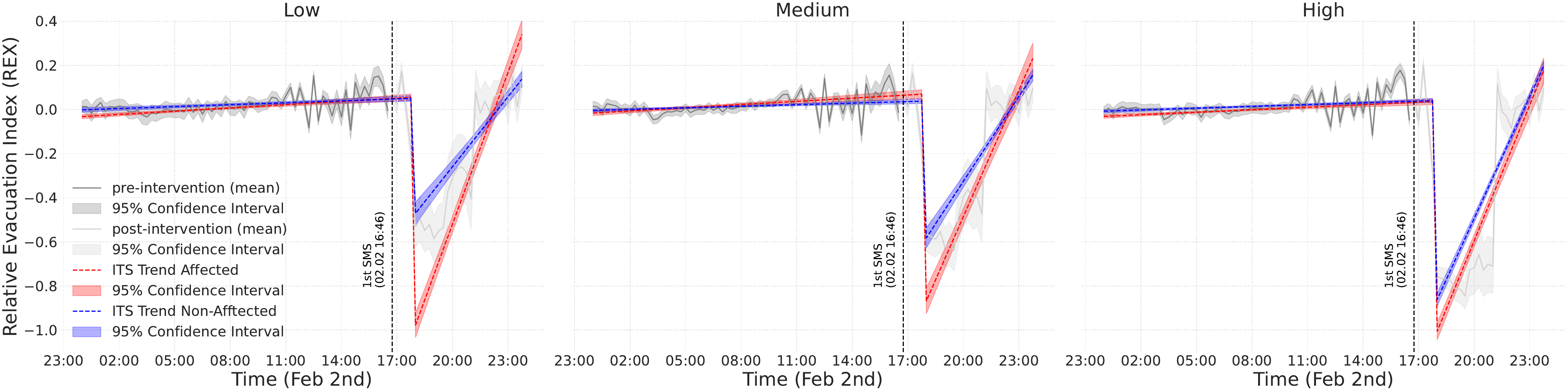}
    \caption{\textbf{Controlled Interrupted Time Series (CITS) stratified by three SEGs (Low (left), Medium (center), High (right) for February 2nd, 2024.} We display the pre-intervention (dark-gray) and post-intervention (light-gray) observed REX. The red line (and red bands) represent the predicted REX trend (and 95\% CI) for the affected group using CITS analysis. The blue line (and blue bands) represent the predicted REX trend (and 95\% CI) for the control group using CITS analysis.}
    \label{fig:model}
\end{figure}

Our results revealed three main findings regarding socioeconomic disparities in evacuation behavior (Figure \ref{fig:model}). First, higher socioeconomic groups demonstrated heightened immediate responsiveness to warning messages, as evidenced by larger negative intervention coefficients ($I_0$). Second, lower socioeconomic groups exhibited stronger differential effects between treatment and control conditions, manifested in both immediate response ($GI_0$) and temporal trends ($GT_{I_0}$). Third, while higher socioeconomic groups showed more rapid recovery patterns ($T_{I_0}$), lower socioeconomic groups in affected areas demonstrated accelerated differential return rates ($GT_{I_0}$). These findings suggest complex socioeconomic heterogeneity in both evacuation decision-making and subsequent return behavior.

\section{Discussion}
\label{sec:discussion}

Our near-real-time (one-day delay) connectivity analysis of the Valpara\'iso wildfires provides quantitative assessments addressing research questions on timing effects, socioeconomic disparities, and spatial consequences of emergency alerts. By integrating high-resolution data with a structured evaluation framework, we offer empirical validation of evacuation behavior and emergency communication effectiveness, extending prior research on disaster response.

Our first finding suggests potential alert fatigue. The initial evacuation alert triggered an immediate response, with connectivity dropping by over 80\% in some areas within 1 hour and 30 minutes, while subsequent alerts showed diminishing effects. Repeated exposure to frequent alerts may reduce responsiveness to actual threats, as previously observed during the Los Angeles wildfires \cite{guardian2025wildfires, ap2025firealert}. These observations contribute to the existing literature by providing near-real-time quantitative assessments of evacuation warnings, as erosion of trust in emergency messaging can lead to delays in life-saving evacuations \cite{mileti1990communication}. Integrating these insights into evacuation alert administration may minimize redundant messages while maintaining public confidence in emergency communication systems.

The second key finding highlights socioeconomic disparities in evacuation patterns \cite{takahiro_yabe_effects_2020, yabe_understanding_2020}. Lower socioeconomic groups (SEGs) were more impacted by the evacuation alert, taking more time not only to evacuate but also to return to normality \cite{thompson_evacuation_2017, burke_exposures_2022}. We observed that high-income SEGs evacuated and returned to normal activity faster, suggesting that populations in wealthier regions experience fewer barriers to evacuation and recovery, likely due to alternative housing options. Given these patterns, communication systems could integrate return-safety notifications to prevent premature re-entry into hazardous areas \cite{burke_exposures_2022, yabe_behavior-based_2023}.

The third key finding concerns evacuation in non-warned areas. High-SEG populations showed less differentiation between warned and non-warned areas, indicating a tendency for precautionary evacuation even without official directives. In contrast, lower-income populations exhibited a stronger differential response between warned and non-warned areas, suggesting a greater reliance on government-issued alerts for evacuation decisions \cite{thompson_evacuation_2017, cursino2024valparaiso}. In non-warned areas, we measured connectivity reductions of -0.47 for low-SEG, -0.56 for medium-SEG, and -0.81 for high-SEG, indicating varied levels of evacuation based on perceived risk \cite{mcconnell2024rare, burke_exposures_2022}. While voluntary evacuations can reduce fatalities, they can also increase transportation congestion and strain emergency response infrastructure \cite{park_post-disaster_2024}.

The spatial diffusion of evacuation behavior beyond targeted areas presents challenges for geographic precision in alert dissemination and resource allocation, a concern previously noted in wildfire evacuation studies \cite{wang_estimating_2014}. Our work contributes to the understanding of risk communication theory and digital crisis response mechanisms by quantifying voluntary evacuations as they occur.

Despite its strengths, this study has methodological limitations. Reliance on mobile phone data from a single carrier, which represents 28\% of the market, may under-represent certain population segments \cite{telefonica_consolidated_2024}. Additionally, socioeconomic classification based on census data may obscure within-zone heterogeneity, limiting our analysis granularity. Attribution of behavioral changes solely to evacuation alerts presents another challenge, as evacuation decisions are influenced by multiple factors, including media coverage, social networks, and direct observation of fire conditions \cite{mcconnell2024rare}. While our CITS methodology mitigates some concerns by comparing affected and non-affected areas, it cannot fully isolate the effect of evacuation alerts from other sources. Furthermore, our REX measure captures relative population movement (in the sense that if there is less activity in a given tower, this means that devices that were once there, are not anymore) but does not distinguish between permanent evacuation, temporary relocation, or routine mobility, potentially affecting the interpretation of post-alert movement patterns.

Future research should explore multi-channel emergency communication, comparing the effectiveness of evacuation alerts, social media updates, and emergency broadcast systems to determine optimal strategies for different socioeconomic groups \cite{federal_communications_commission_wireless_2024, senapred2024alerts}. Longitudinal studies tracking alert responsiveness across multiple disasters could identify desensitization patterns and develop countermeasures to maintain public engagement with warnings. Research should also examine how linguistic framing, information sequencing, and geographic specificity influence evacuation compliance \cite{mejova2019effect}. Integrating near-behavioral data with survey methods could deepen our understanding of how individuals interpret and act upon emergency messages. Computational modeling using mobile network data could refine evacuation predictions and improve emergency response simulations, enabling authorities to optimize alert strategies before future crises \cite{yabe2022mobile}.

Our findings reveal insights for disaster management that may help refine evacuation protocols to implement multi-tiered, phased evacuations. Prior research, primarily based on interviews and surveys, has evidenced similar patterns of socioeconomic evacuation inequalities, particularly regarding transportation access and trust in authorities \cite{thompson_evacuation_2017}. We contribute to current literature by providing quantification of evacuation patterns as they occur, illuminating temporal and socioeconomic evacuation dynamics that complement survey-based approaches. These insights may help prioritize at-risk communities first, preventing bottlenecks and reducing transportation congestion \cite{yabe_understanding_2020}. Moreover, these refinements may enhance evacuation compliance, reduce infrastructure strain, and improve resource allocation during crises.

With climate change increasing wildfire severity, optimizing emergency communication systems is essential for population safety. The observed over-evacuation in non-targeted areas highlights the need for precision in alert dissemination to prevent infrastructure congestion \cite{mcconnell2024rare}. Our findings contribute to evidence supporting the benefits of integrating mobile connectivity patterns into disaster management—a critical adaptation pathway for societies facing increasing climate-driven hazards.

\section{Methods}
\label{sec:methods}
\footnotesize
\paragraph{Identifying warned towers.} The messages were provided to us in an Excel sheet in the following form: \texttt{date}, \texttt{hour}, \texttt{event}, \texttt{threat}, \texttt{message}, \texttt{polygon}, \texttt{region}, \texttt{X}. The \texttt{date} and \texttt{hour} fields were two separate fields in the format DD/MM/YYYY and HH:MM, respectively, the latter in the 24-hour format. The first message was time-stamped with \texttt{date} \texttt{2/2/2024} and \texttt{time} \texttt{16:45}. The \texttt{event} was of type \texttt{Evacuaci\'on (Incendio Forestal)} (tr. Evacuation (Forest Fire)). \texttt{threat} was \texttt{Incendio Forestal} (tr. Forest Fire). The \texttt{message} was \texttt{SENAPRED: Por incendio forestal evacue sector Quebrada Escobares, comuna de Villa Alemana} (tr. SENAPRED: Due to a forest fire, evacuate the Quebrada Escobares sector, Villa Alemana commune). \texttt{polygon} was \texttt{Comuna de Villa Alemana} (tr. Commune of Villa Alemana); \texttt{region} was the affected region, in the case of the first message, it was \texttt{Valpara\'iso} and finally \texttt{X} was the X/Twitter message sent. In the first message, this was \texttt{https://twitter.com/Senapred/status/1753503622174761190}. Unfortunately, the information did not include a ``real" polygon of any affected areas, according to the evacuation alert. There was only a general area to which the language alluded. In the above message, this was Quebrada Escobares, east of the town of Quilpu\'e. To formalize this and provide a more specific area, we took all the unique places mentioned in all alerts that were sent and used the Nominatim geocoding service\footnote{\url{https://nominatim.org/}} to retrieve the latitude and longitude coordinates for each place. These identified places are the green dots in Figure \ref{fig:evacuation_analysis}B. The 5 km of the affected towers were identified as originating from these points. All towers that were not warned were used as controls; see the study design below.

\paragraph{Measuring connectivity.}
We analyzed anonymized eXtended Detail Records (or XDRs) of a leading mobile phone operator in South America, Telefónica Movistar, which had a market share of 28\% in June 2024~\cite{telefonica_consolidated_2024}. The data set contains the events of all the devices in the network. We have shown in other works that this is highly correlated with census information in terms of population distributions \cite{gauvin_gender_2020,elejalde_social_2024}. XDRs capture detailed metadata on mobile network activities, offering temporally granular insights into user behaviors. An XDR is a tuple \((n, t, A, k)\); unlike more common CDRs, XDRs involve only one tower \(A\). Here, \(n\) is the caller's identifier (ID), \(t\) is the timestamp of record creation, and \(k\) is the amount of data downloaded (in kilobytes) \cite{pappalardo_evaluation_2021}. Records were aggregated at 15-minute intervals by counting distinct anonymized phone IDs per tower, with timestamps rounded down to the nearest 15-minute block: this is what we call ``connectivity patterns''. Notice that there is one way in which this is akin to a simple mobility measure: IDs that are no longer present in one tower are supposed to have moved to another, unknown one. We established two comparison periods: a \textbf{baseline days} (2024-01-25 07:00 to 2024-01-28 00:00) and a \textbf{fire week} (2024-02-01 07:00 to 2024-02-04 00:00). The baseline period comprised $N = 57,140,040$ total records, with a mean of $\bar{x} = 575,094$ unique phone IDs ($\sigma = 16,213$) distributed across $\bar{x} = 584$ towers ($\sigma = 1$). The wildfire period showed comparable metrics, with $N = 62,579,019$ total records, $\bar{x} = 580,758$ unique phone IDs ($\sigma = 16,219$) distributed across $\bar{x} = 583$ towers ($\sigma = 3$). This baseline served as a reference for normal tower usage patterns, enabling direct comparison with connectivity patterns observed during the wildfires.

\paragraph{Assigning socio-economic groups.} Due to the lack of reliable self-reported socioeconomic data, we estimated a tower's socioeconomic status based on its location. We used 2017 census data from the National Institute of Statistics of Chile to classify the location of the towers into three socioeconomic categories: Low, Middle, and High. To estimate socioeconomic status, we first assigned it to the entire census zone, using the proportion of individuals with higher education as an indicator. We also assessed the correlation between educational level and the Socio-Material Territorial Index\footnote{More detailed information about this index is available at: \href{https://ideocuc-ocuc.hub.arcgis.com/maps/c83a1ea2c31b4850b65a481b21e4919f/about}{this link} (in Spanish).}, which yielded a high correlation coefficient of $\rho$ = 0.93~\cite{naushirvanov_evacuation_2024}. We decided to use only educational level, as the methodology behind the Socio-Material Territorial Index is less interpretable. Once we had the socioeconomic index, we divided census zones into socioeconomic categories using population quantiles. This ensured a balanced representation across groups and maintained adequate sample sizes for each category. After determining the socioeconomic status for all census zones, we assigned these zones' status to the communication towers located within them. Each census zone, on average, contains 1.53 telecommunication towers and has a total population of approximately 2,436 individuals.
This classification approach allowed us to identify key behavioral trends and differences between socioeconomic groups while maintaining a balance and avoiding overly detailed subdivisions that might obscure meaningful patterns.

\paragraph{Study design and statistical analysis.}
We employed a quasi-experimental design using a controlled interrupted time series (CITS) to assess behavioral responses to wildfire warning messages \cite{mcdowall_interrupted_2003,lopez_bernal_use_2018}. For this study, we defined the control group based on geolocation data, selecting cellular towers that were never directly addressed by the warning SMSs and, therefore, should have remained unaffected by the wildfires. Although individuals connected to these non-warned towers may have been exposed to warning messages, their behavioral response is expected to be primarily preventive. To account for response latency, we incorporated a delayed intervention effect, informed by previous studies on emergency evacuations, which indicate reaction delays due to factors such as health status, social influences, and affiliations \cite{fahy_toward_2001,liu_analysis_2020}. Model calibration across a range of delays identified 1 hour and 15 minutes as the optimal lag for capturing observed evacuation patterns (Supplementary Figure \ref{fig:progr_delay_model_comp}). We estimated the intervention effect using the following CITS model:
\begin{equation}
\label{eq:CITS_model}
    Y = \beta_0 + \beta_1 T + \beta_2 I_0 + \beta_3 T_{I_0} + 
        \beta_4 G + \beta_5 GT + \beta_6 GI_0 + \beta_7 GT_{I_0} + 
        \epsilon
\end{equation}
\noindent where \(Y\) represents the predicted REX, \(T\) denotes time (measured in 15-minute intervals), \(I_0\) is a post-event indicator set to 1 after the intervention (incorporating the delay), and \(T_{I_0}\) captures the trend change following the intervention. \(G\) distinguishes between affected and control groups. The Newey-West variance estimator was used to correct for heteroskedasticity and autocorrelation, ensuring robust inference. For completeness, we also include in the Supplementary Materials the results from the uncontrolled Interrupted Time Series analysis for the entire observed period (Table \ref{tab:regression-ITC} and Figure \ref{fig:model-ITS}).

\normalsize

\section*{Abbreviations}
\begin{itemize}
    \item \textbf{BIC} - Bayesian Information Criterion
    \item \textbf{CDR} - Call Detail Records
    \item \textbf{CI} - Confidence Interval
    \item \textbf{CITS} - Controlled Interrupted Time Series
    \item \textbf{CLST} - Chile Standard Time
    \item \textbf{FCC} - Federal Communications Commission
    \item \textbf{ITS} - Interrupted Time Series
    \item \textbf{ITU} - International Telecommunication Union
    \item \textbf{REX} - Relative Evacuation Index
    \item \textbf{SAE} - Sistema de Alerta de Emergencia (Emergency Alert System in Chile)
    \item \textbf{SEG} - Socio-Economic Group
    \item \textbf{SMS} - Short Message Service
    \item \textbf{WEA} - Wireless Emergency Alerts
    \item \textbf{XDR} - eXtended Detail Records
\end{itemize}

\subsection*{Funding}

LF thanks Telefónica Chile for funding and support. This research was supported by FONDECYT Grant N°1221315 to LF. LF, KK acknowledge support from the Lagrange Project of ISI Foundation funded by CRT Foundation. MK acknowledges funding from the National Laboratory for Health Security (RRF-2.3.1-21-2022-00006), the ANR project DATAREDUX (ANR-19-CE46-0008); the SoBigData++ H2020-871042; the MOMA WWTF; and the COLINE DUT-FFG projects.

\subsection*{Authors' contributions}

\noindent \textbf{Conceptualisation}: all authors; 
\textbf{Methodology}: all authors; 
\textbf{Formal analysis}: EE;
\textbf{Investigation}: EE, LF; 
\textbf{Writing -- Original Draft}: LF, EE, KK; 
\textbf{Writing -- Review \& Editing}: all authors.

\bibliographystyle{abbrv}
\bibliography{arxiv}

\clearpage
\section*{Supplementary Materials}

\setcounter{table}{0}
\renewcommand{\thetable}{S\arabic{table}}
\setcounter{figure}{0}
\renewcommand{\thefigure}{S\arabic{figure}}

\begin{table}[htbp]
\small
    \centering
\begin{tabular}{ccccc}
\toprule\toprule
    ID & Datetime & High & Medium & Low \\
    \midrule
    0 & 2024-02-02 16:45 & 2 & 0 & 0 \\
    1 & 2024-02-02 18:30 & 12 & 12 & 4 \\
    2 & 2024-02-02 18:45 & 4 & 8 & 3 \\
    3 & 2024-02-02 19:45 & 0 & 7 & 6 \\
    4 & 2024-02-02 21:15 & 1 & 1 & 0 \\
    5 & 2024-02-03 10:45 & 1 & 0 & 1 \\
    6 & 2024-02-03 12:15 & 4 & 1 & 3 \\
    7 & 2024-02-03 13:00 & 1 & 2 & 0 \\
    8 & 2024-02-03 14:15 & 6 & 6 & 2 \\
    9 & 2024-02-03 14:30 & 0 & 1 & 1 \\
    10 & 2024-02-03 15:00 & 17 & 0 & 0 \\
    11 & 2024-02-03 17:00 & 0 & 1 & 0 \\
    12 & 2024-02-03 18:00 & 2 & 0 & 6 \\
    13 & 2024-02-03 18:30 & 0 & 0 & 2 \\
    \midrule
     &  & \textbf{50} & \textbf{39} & \textbf{27} \\
    \bottomrule\bottomrule
\end{tabular}
    \caption{Time of the SMSs that included new affected towers in the warning message. The data shows the number of new towers warned by each SMS stratified by socio-economic groups.}
    \label{tab:message_times}
\end{table}

\begin{figure}[htbp]
    \centering
    \includegraphics[width=.77\linewidth]{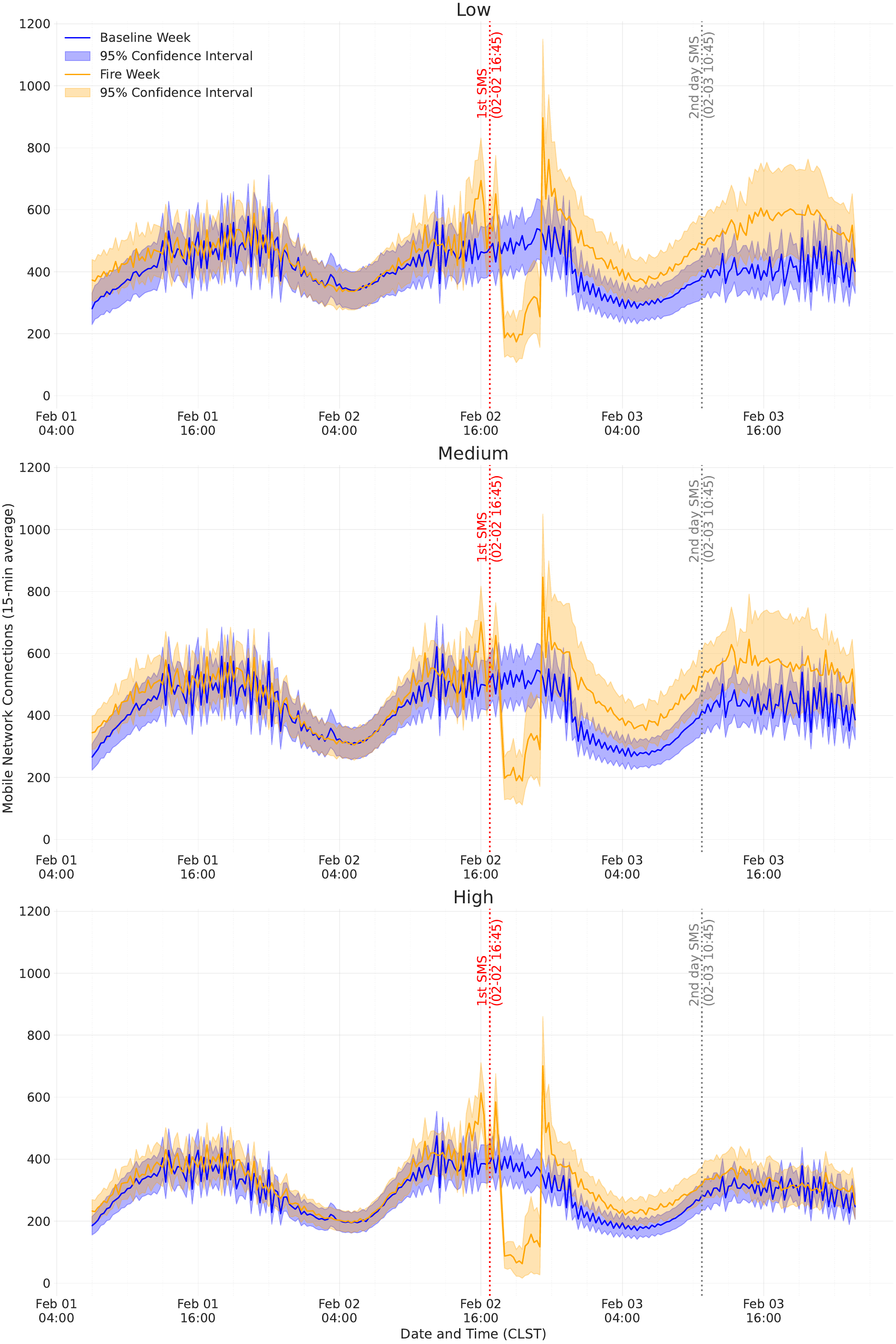}
    \caption{\textbf{Mobile phone tower connectivity patterns during a wildfire evacuation event by socio-economic group: low (top), medium (middle) and high (bottom)}.}
    \label{suptbl:supp_rawsocioecon}
\end{figure}

\begin{figure}[htbp]
    \centering
    \includegraphics[width=1\linewidth]{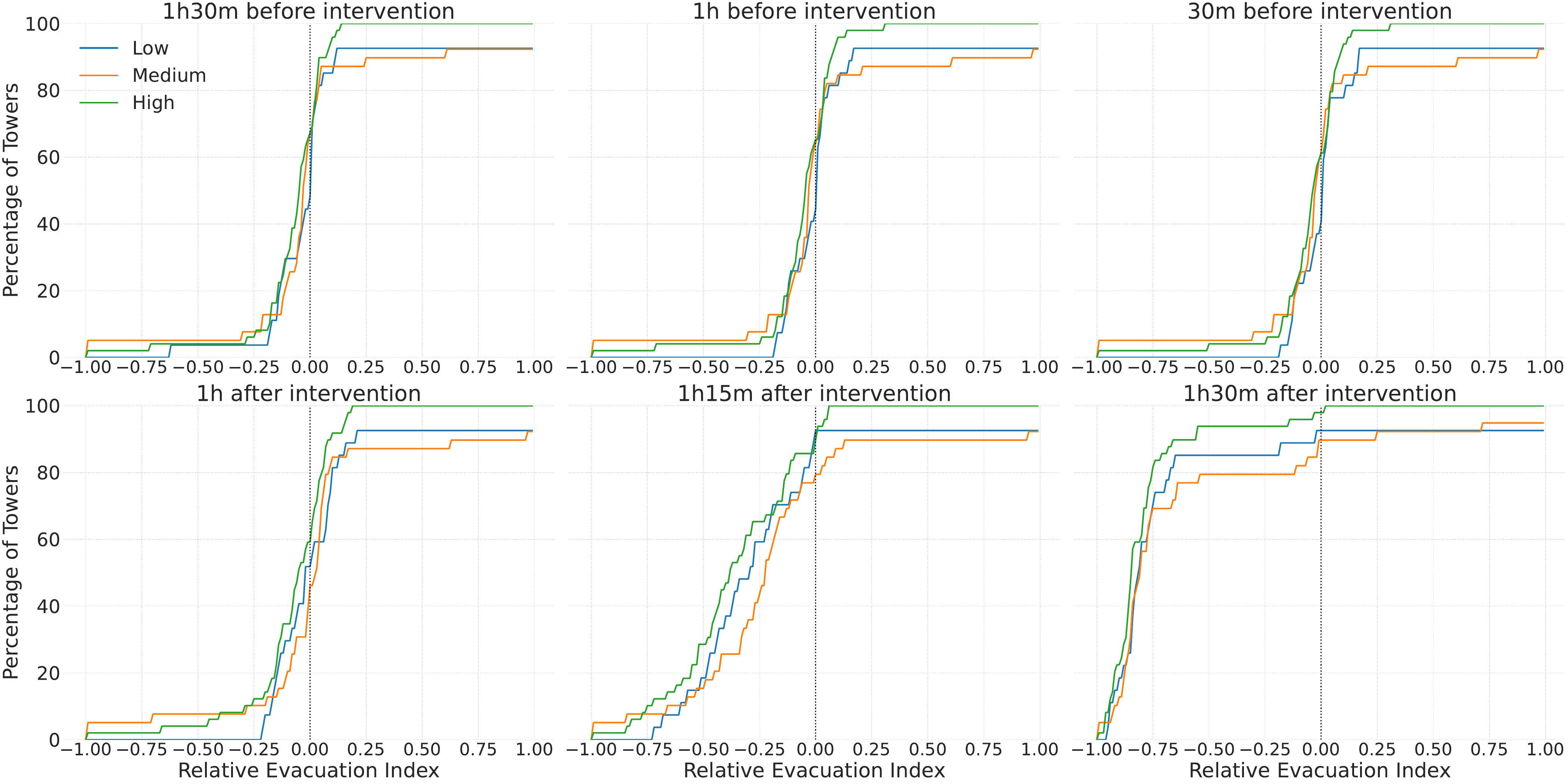}
    \caption{\textbf{Cumulative distribution functions (CDFs) of tower-level REX, stratified by socioeconomic status}. The figure presents six time-sliced CDFs capturing population displacement patterns before and after the intervention across low (blue), medium (orange), and high (green) socioeconomic areas. Pre-intervention panels show behavior at 1.5 hours, 1 hour, and 0.5 hours (30 minutes), while post-intervention panels display changes at 1 hour, 1 hour and 15 minutes, and 1 hour and 30 minutes. The x-axis represents the observed REX, where 0 indicates behavior consistent with the baseline days, negative values indicate evacuation relative to baseline, and positive values represent more connections relative to baseline. The y-axis shows the cumulative percentage of towers exhibiting REX lower than or equal to the corresponding x-value up to this time.}
    \label{fig:CDFs_diff_first_ever_norm}
\end{figure}

\begin{figure}[htbp]
    \centering
    \includegraphics[width=\linewidth]{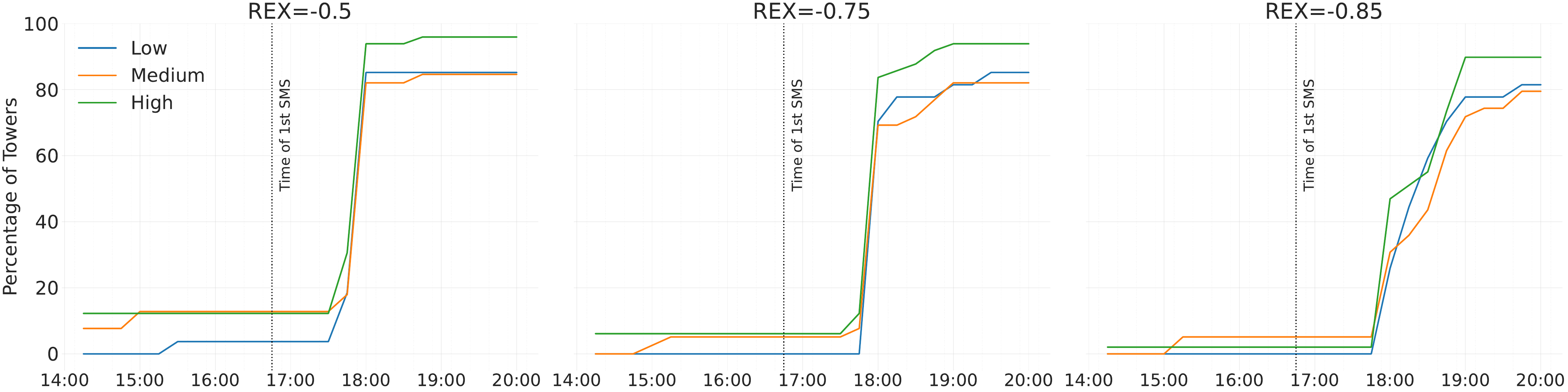}
    \caption{\textbf{Cumulative distribution function of tower REX over time}. The x-axis represents time, while the y-axis indicates the percentage of towers that reached the specified REX rates (-0.50, -0.75, and -0.85). The lines are stratified by socio-economic groups (SEGs): low (blue), medium (orange), and high (green). The dashed line marks the time at which the first evacuation SMS was sent.}
    \label{fig:CDFs_time_first_ever_norm}
\end{figure}

\begin{figure}[htbp]
    \centering
    \includegraphics[width=1\linewidth]{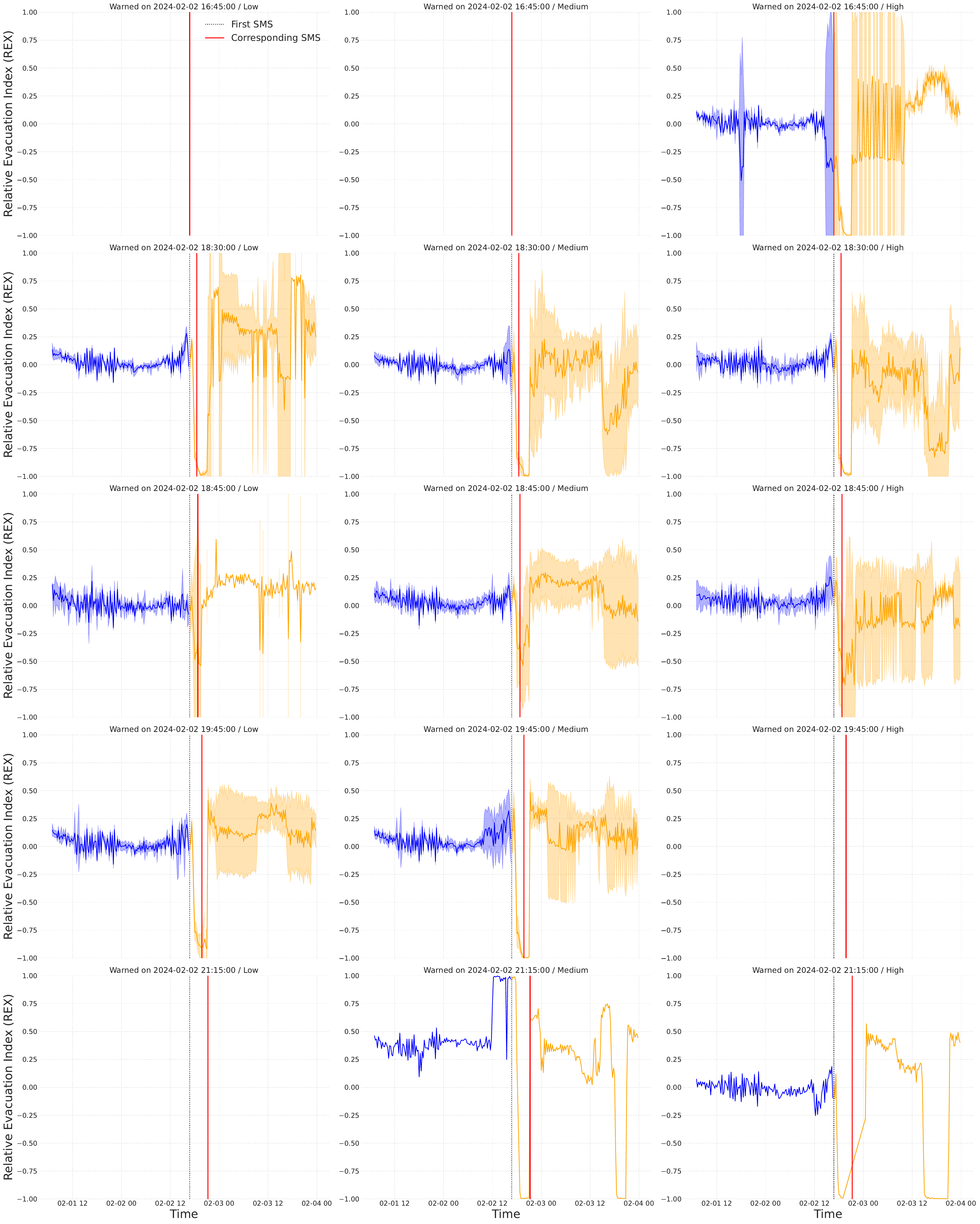}
    \caption{\textbf{REX for affected towers notified on the first day of the wildfires.} Every row represents the evacuation behavior for affected towers that were notified for the first time in the corresponding warning SMS. Columns represent different Socio-Economic Groups (SEG).}
    \label{fig:rex-first-day}
\end{figure}

\begin{figure}[htbp]
    \centering
    \includegraphics[width=0.6\linewidth]{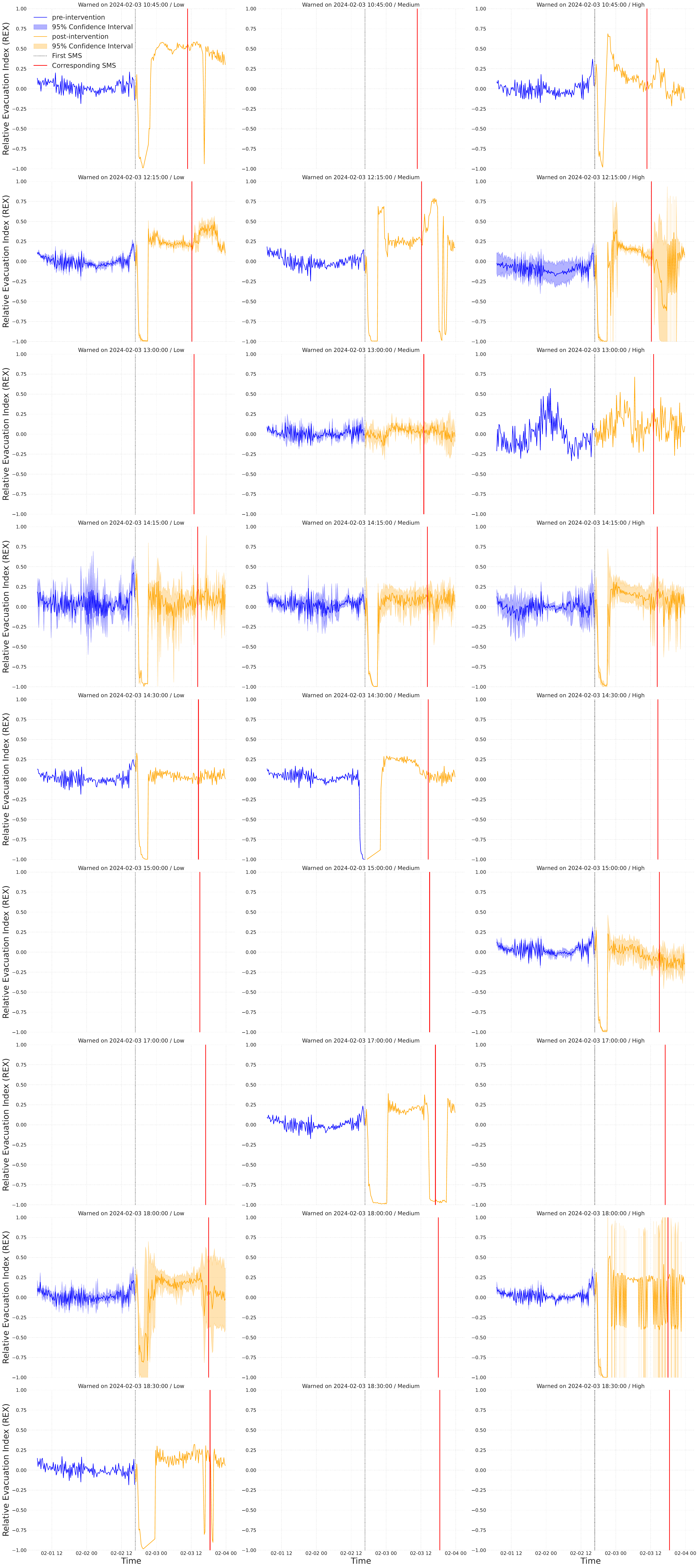}
    \caption{\textbf{REX for affected towers notified on the second day of the wildfires.} Every row represents the evacuation behavior for affected towers that were notified for the first time in the corresponding warning SMS. Columns represent different Socio-Economic Groups (SEG).}
    \label{fig:rex-second-day}
\end{figure}

\begin{figure}[htbp]
    \centering
    \includegraphics[width=1\linewidth]{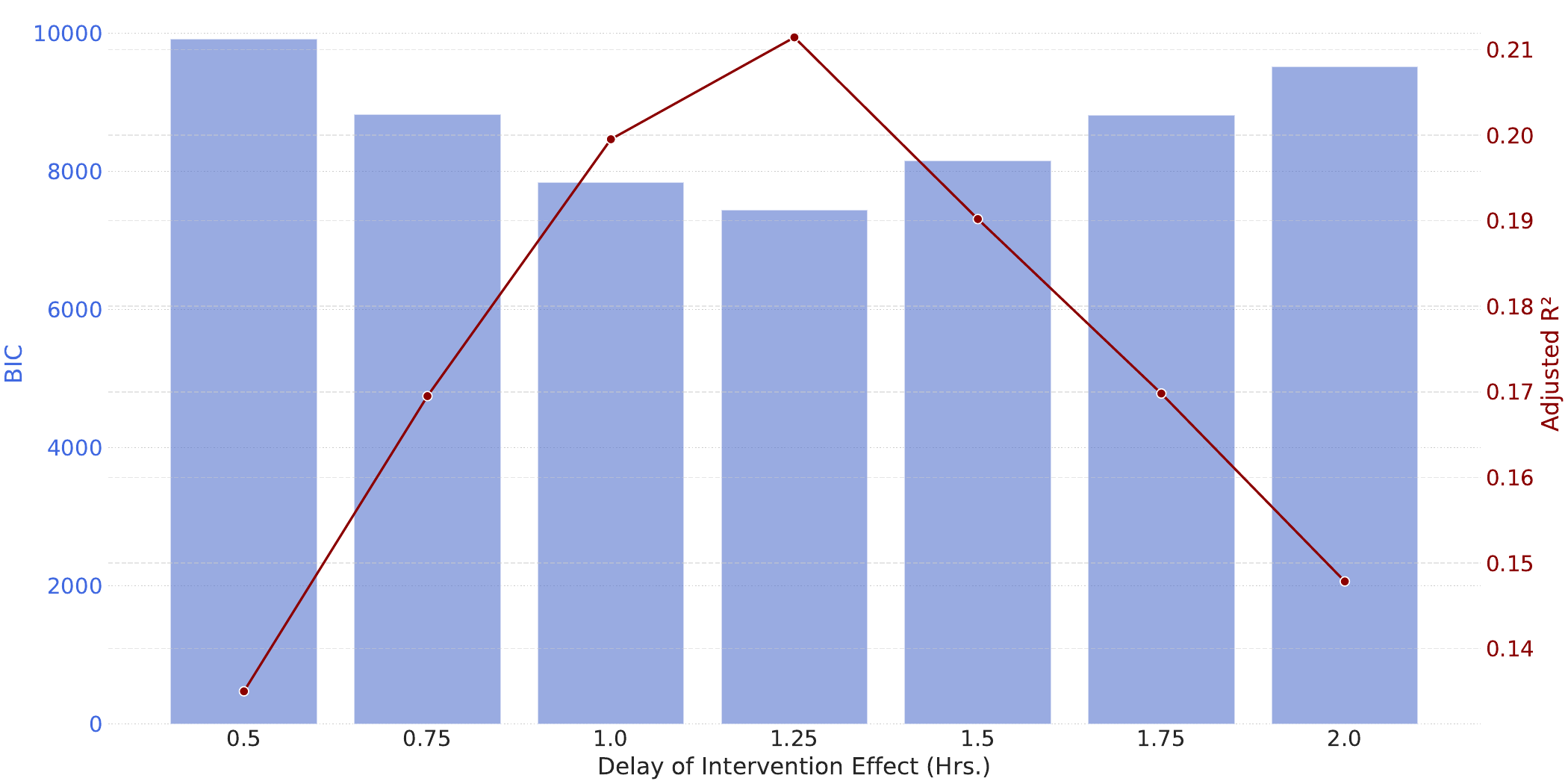}
    \caption{Comparison between Interrupted Time Series (ITS) models for affected towers with different delays. The x-axis represents the range of delays tested in 15-minute steps (from 30 minutes (0.5 Hrs.) to 120 minutes (2.0 Hrs.)). The bars represent the BIC for the regression using the corresponding delay (left y-axis). The line plot represents the Adjusted $R^2$ for the regression using the corresponding delay (right y-axis).}
    \label{fig:progr_delay_model_comp}
\end{figure}

\begin{figure}[htbp]
    \centering
    \includegraphics[width=1\linewidth]{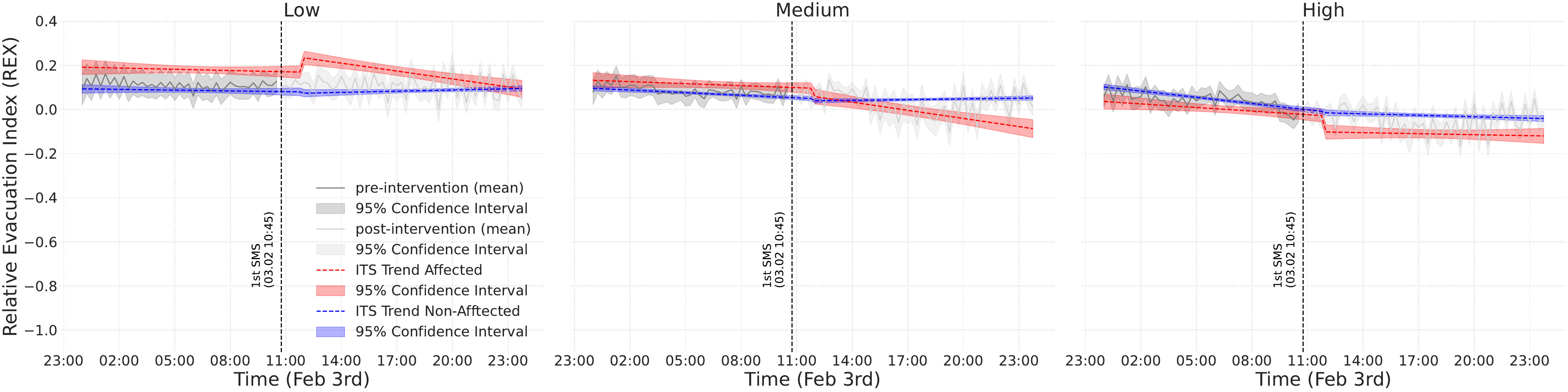}
    \caption{\textbf{Controlled Interrupted Time Series (CITS) stratified by three socio-economic groups (Low (left), Medium (center), High (right) for February 3rd.} The data displays the pre-intervention (dark-gray) and post-intervention (light-gray) observed REX. The red line (and red bands) represent the predicted REX trend (and 95\% CI) for the affected group using CITS analysis. The blue line (and blue bands) represent the predicted REX trend (and 95\% CI) for the control group using CITS analysis.}
    \label{fig:model_day2}
\end{figure}

\begin{figure}[htbp]
    \centering
    \includegraphics[width=1\linewidth]{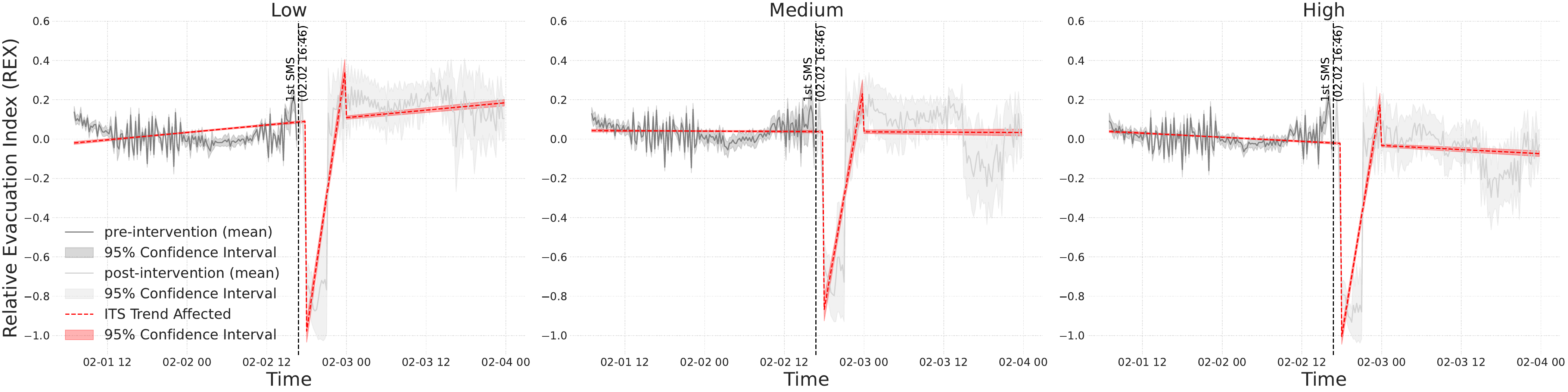}
    \caption{\textbf{(Uncontrolled) Interrupted Time Series (ITS) stratified by three socio-economic groups (Low (left), Medium (center), High (right) from February 1-3.} The data displays the pre-intervention (dark-gray) and post-intervention (light-gray) observed REX. The red line (and red bands) represent the predicted REX trend (and 95\% CI) for the affected group using ITS analysis.}
    \label{fig:model-ITS}
\end{figure}

\begin{table}[htbp]
\centering\scriptsize
\begin{tabular}{lccc}
\toprule\toprule
& \multicolumn{3}{c}{Socio-Economic Groups}\\
\cmidrule(lr){2-4}
& Low & Medium & High \\
\midrule
Intercept & -0.02*** & 0.043*** & 0.039*** \\
          & (-0.027--0.012) & (0.036-0.05) & (0.033-0.045) \\
$T$ & 0.001*** & 0.0 & 0.0*** \\
         & (0.001-0.001) & (0.0-0.0) & (-0.001--0.0) \\
$I_0$ & -0.786*** & -0.666*** & -0.724*** \\
 & (-0.829--0.742) & (-0.71--0.623) & (-0.755--0.694) \\
$T_{I_0}$ & 0.057*** & 0.048*** & 0.052*** \\
 & (0.052-0.061) & (0.043-0.052) & (0.048-0.055) \\
\midrule\midrule
\multicolumn{3}{@{}l@{}}{\footnotesize Note: $^{**}p<0.01$; $^{***}p<0.001$}
\end{tabular}
\caption{Interrupted Time Series (ITS) Analysis (uncontrolled) for different Socio-Economic Groups (SEG).}
\label{tab:regression-ITC}
\end{table}

\end{document}